\documentclass[]{hdsr}
\begin{document}
\newgeometry{top=0.9in, bottom=1in,right=1in,left=1in}
  \title{Measuring Geometric Similarity Across Possible Plans for Automated Redistricting}
  \author{Gilvir Gill}
  \maketitle
\begin{center}
  \vspace{-0.825in}
  \begin{tabular}{cc}
    Gilvir Gill
   \\[0.20ex]
   {Stony Brook University} \\
  \end{tabular}
\end{center}
\vspace{-0.2in}
\vspace{0.2in}

\section{Introduction}
Congressional districts in the United States are typically drawn following the decennial census
by state legislatures or independent commissions, and the process of redistricting
can have a major effect on electoral outcomes. As most states will be required to redraw their districts
in 2021, gerrymandering, the politically or racially motivated manipulation of districting plans, has become an
increasingly relevant issue.
Since \citet{chen2013}, common techniques for exposing gerrymandering have used computers to generate 
large numbers of random district plans, comparing an enacted plan to the set of all 
possible ones.
Among these, randomized region growing
and recursive cluster merging (\citet{chen2015}) have historically been popular.
With \citet{mattingly2014redistricting}, the new gold standard has become
using Markov chain Monte Carlo (MCMC) to draw random samples from a distribution. Additional
innovations have been made on MCMC approaches for redistricting analysis, most notably the spanning-tree recombination
approach from \citet{deford2019recombination}. Other approaches, such as the one used in this paper, 
modify these sampling approaches to bias transitions towards higher-scoring
solutions by introducing an objective function that quantifies legal and political 
guidelines for redistricting.
However, due to randomization shortcomings and lack of
deep comparison of these different techniques, automated redistricting and sampling approaches remain
open problems. And while previous work such as \citet{deford2019recombination} tends to compare groups of 
redistricting plans by outcome (for example, distributions of minority populations through each of the final districts,
Polsby-Popper compactness, or projected electoral results), there is still a need for methods to compare plans 
purely on their geometry.

This paper attempts to make some progress in allowing comparisons between different distributions
and sampling techniques, by presenting a graph-topological measure for 
comparing the geometry of any given pair of congressional plans. 
This measure has an intuitive and interpretable definition,
being equal to the percentage of the state's area or population that remain
in the same congressional district across the two plans, 
and is shown to be computationally efficient and therefore 
feasible to calculate for a large number of redistricting plans.

\section{Objective Functions and Measures for Redistricting}

An objective function approximates how well a given districting plan 
satisfies state-specific constraints and gives a simple mechanism to adapt to differing
legal doctrine on redistricting across states. 
For instance, in compliance with state and federal guidelines, 
commissions consider racial demographics in the process to ensure fair representation, though
recent studies challenge that approach (\citet{chen_stephanopoulos_2020}).
Additionally, states will often require, formally or informally, that new
redistricting plans be similar to the currently enacted plans. This could either
require commissions to qualitatively consider a geometric similarity in the plans,
or could involve requirements such as minimizing the number of people who are moved
to a new district between two plans. In both of these cases, an old plan and a new
enacted plan ought to allocate electoral precincts to districts to maximize the 
intersection of some precinct-level quality, usually area or population.

However, there is a lack of clear guidelines on
what similarity means in the context of two sets of plans, whereas concepts such as compactness
and partisan fairness have well-defined measures (Polsby-Popper scores and efficiency gap, respectively). 

We address the lack of a robust mathematical formulation for cross-plan similarity by defining
a new measure and corresponding algorithm that makes use of bipartite maximum-weight matching, and present an example use-case. 
Specifically, we consider how the geometry and demographics of a state determine the pair-wise similarity 
between two plans drawn from the optimal recombination approach. For example, states such as Texas and North Carolina
are topologically thin in their
westernmost sections, and any compact districts containing El Paso and Western North Carolina are largely predetermined
compared to any one part of Georgia, where no section of the state tends to extend out. This, in turn, could increase
the expected similarity between two arbitrary plans.

\section{Methodology}

We propose the following measure for similarity in districts: Let $I(a, b)$ be the 
sum of the area of the electoral precincts (or a smaller unit, such as census blocks)
shared by districts $a$ and $b$ from separate districting plans. 
For two given plans, with not necessarily the same number of districts,
$D_1 = \{ d_1^1, d_1^2, \dots d_1^n\}$ and $D_2 = \{ d_2^1, d_2^2, \dots d_2^k \}$ (where $n \leq k$; the number of districts in each plan
need not be the same), reorder the indices for $D_2$ to minimize the following function, where $A$ is the
total area of the state:

\begin{align*}
    f(D_1, D_2) &= \frac{1}{A} \sum_{i=1}^{n} I(d_1^i,d_2^i)
\end{align*}

Considering the minimum value of $f$ across all possible pairings of districts in $D_1$ and $D_2$ is akin to finding the mapping from old district numbers to new district numbers such that
the largest portion of the state's area will stay in the same district between two different plans. If the redistricting plans are
not renumbered, the similarity measure becomes effectively useless; one randomly sampled plan may have 
District 1 assigned to the northwest region
of a state, for example, while another randomly sampled plan could have District 1 assigned to the southeast.

Alternatively, we can replace $I(a,b)$ with a  
population intersection function to get the district number mapping that keeps the most
people possible in their current congressional district given a set of equal-population boundaries. This would
likely be a more interesting measure on states such as New York, where a majority of the state's population resides 
in a small region of the state, and would be more in line with informal guidelines that incumbents keep
the largest possible chunk of their constituents between redistrictings.

We can find these best mappings in $\Theta(n + m^\frac{5}{2})$, where $m$ is the larger number of districts and $n$ the total number of
electoral precincts in the state, by adding each precinct's 
area to entries in a matrix for each intersection and then using a bipartite 
maximum-weight matching algorithm (see \citet{hopcroft1973} or newer). Note that
this problem is sometimes described as a linear sum assignment problem (as in \citet{crouse2016})---finding a boolean cost-matrix $X^*$:

\begin{align*}
    X^* &= \arg \max_X \sum_{i=1}^n \sum_{j=1}^k I(d_1^i, d_2^j) \cdot X_{i,j}
\end{align*}

Where the matrices $X$ have the additional restrictions that
each row is assigned exactly one column, while every column is assigned exactly 
zero or one rows. The resulting matrix $X^*$ therefore has the property that 
$X^*_{i,j} = 1$ if and only if districts $d_1^i$ and $d_2^j$ would be assigned the
same district numbers in a naming system that maximizes the portion of the state kept
under the same district number.

This new measure of similarity accounts for the randomness in labeling for district plans, supports comparison between
plans with different numbers of districts (important if the number of districts changes due to shifting demographics),
and only requires polynomial time. 

Maximizing this measure has desirable graphical consequences, in addition to its 
potential for quantifying similarity and incumbency advantage. For example, bipartite
maximum-weight matching is used in the Stony Brook 
PoliTech Automated Redistricting System (ARS) to keep district numbers and colors
consistent between existing plans and algorithmically generated ones (see Figure 1). 

\begin{figure}[H]
    \centering
    \includegraphics[width=6in]{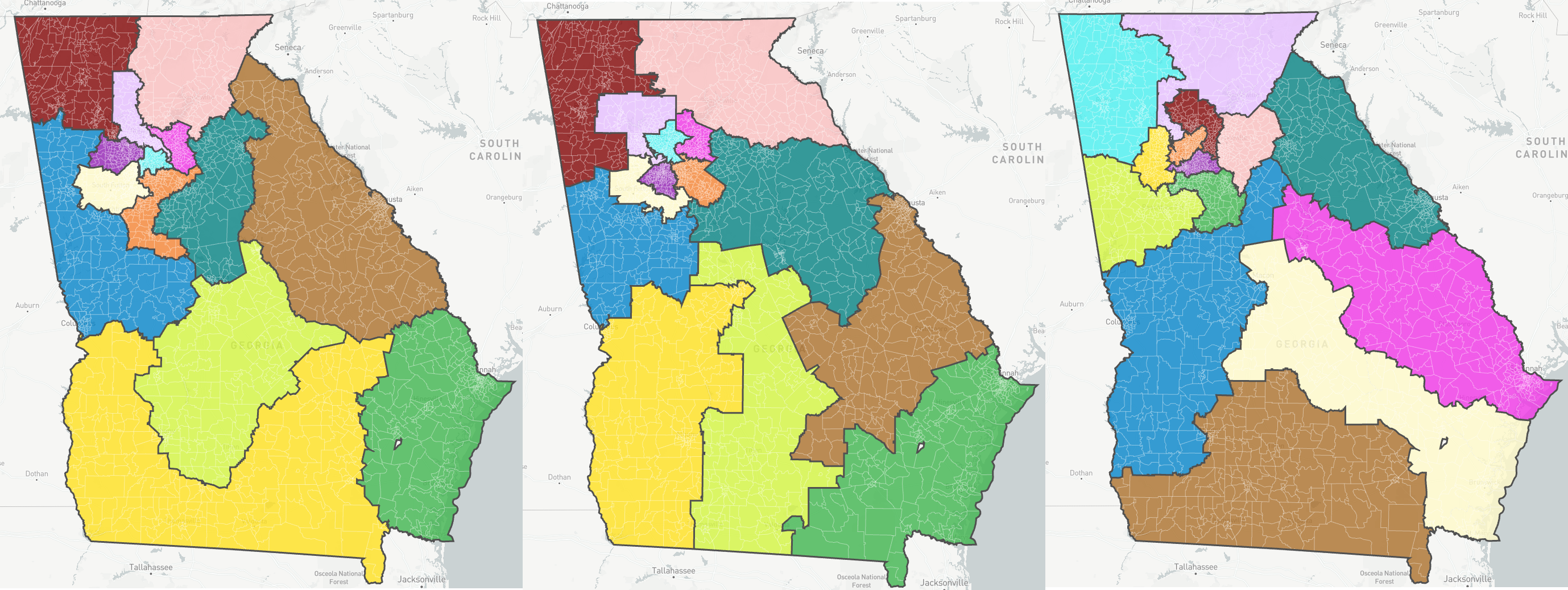} \\
    \caption{Original plan for Georgia (center), generated plans with and without
    measure-based color-matching (left; right, respectively)}
    \vspace{-0.125in}
    \label{fig:my_label}
\end{figure}

In order to highlight a potential use-case for the measure in quantifying similarity,
we generate a set of $N$ districting plans, and 
calculate the similarity score across all $\binom{N}{2}$ pairs of plans.
In this introductory report, we perform our analysis on an a modified version of the MCMC recombination approach that 
attempts to optimize population equality more aggressively by taking the highest scoring transition with respect to a
population-equality objective function, rather than being concerned with staying true to the underlying distribution of
plans that meet a set of requirements. This approach randomly chooses two neighboring districts, merges them into a larger 
cluster, and splits that cluster into two new districts that are more population-equal.
For splitting the clusters, we use the approach from \citet{deford2019recombination}, generating a random spanning-tree 
and taking a cut along the tree to ensure that we split the cluster into two new contiguous districts. 
However, rather than using a random tree cut probabilistically or taking the first cut within a given threshold, 
we check all cuts to find one that best optimizes our population equality. In future analysis, we plan to use this 
measure to compare this modified recombination approach to the traditional MCMC approach. However, since these 
results were focused on observing the differences between states rather than between algorithms, we have left this
analysis out.

We run the algorithm across eight states with differing geometries
and congressional delegation sizes, seeding them with the recursive merging algorithm
from \citet{chen2015}, generating $N=50$ runs in parallel (each with $50 \cdot m$ recombinations, where $m$ is the number of
districts), and computing the similarity scores with the new measure.

\section{Analysis and Results}

\begin{figure}[!htbp]
    \centering
    \includegraphics[width=6in]{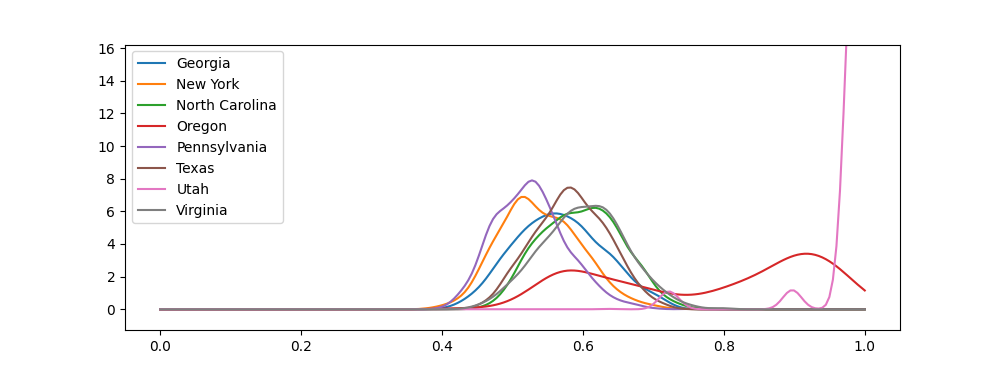} \\
    \caption{Pairwise area similarity scores for generated plans (density function, 1225 pairs)}
    \vspace{-0.125in}
    \label{fig:my_label}
\end{figure}
\begin{figure}[!htbp]
    \centering
    \includegraphics[width=6in]{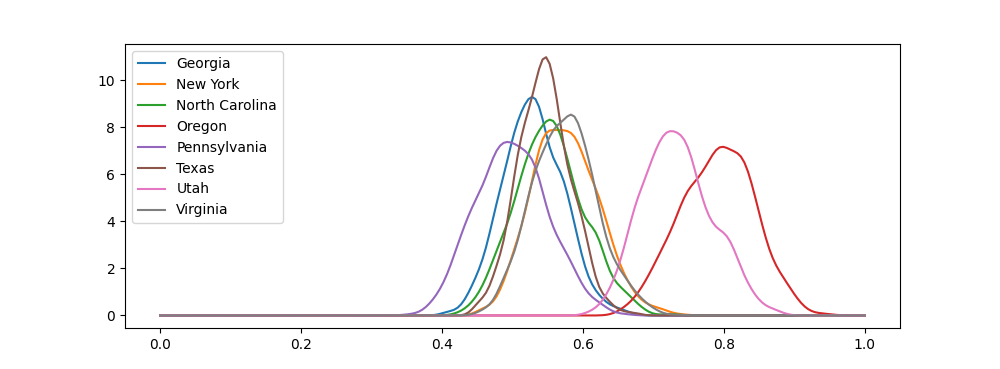} \\
    \caption{Pairwise population similarity scores for generated plans (density function, 1225 pairs)}
    \vspace{-0.125in}
    \label{fig:my_label}
\end{figure}

Running the algorithm with a sample of 1225 pairs per each of the eight states, 
I find that most states have a mean similarity between 0.5 and 0.6. The lack
of similarity scores less than 0.4 is not surprising, largely due to the 
nature of maximal matching; for example, consider that the theoretic minimum
similarity score on the set of all ways to split a circle radially into 
four equal pieces is 0.5.
Oregon and Utah have mean area similarity scores 
of 0.776 and 0.97, respectively, indicating shortcomings of an optimal recombination approach on states with a 
small number of congressional districts compared to a truly random approach.
North Carolina and Texas have higher mean similarities than Georgia and Pennsylvania
(0.598, 0.583; 0.567, 0.526, respectively), showing that our initial theories on the influence of topological bottlenecks
may hold. However, without first quantifying
geometric features of states and normalizing for variables such as population density and clustering,
we cannot draw further conclusions. 
With more resources and time, applying this measure could give meaningful information about the geometric factors that decide the intrinsic similarity between plans. Notably, it is not the specific takeaways from this analysis that are important, but rather, a demonstration 
of how this measure could be useful for understanding inherent biases in both the geometry and graph-topology of a state, and the
behavior of redistricting algorithms at large.

Regardless, disparities between population and area similarity are interesting to note.
For states such as Utah and Oregon, a disproportionate number of electoral precincts
relative to area are found in cities such as Portland and Salt Lake City, which make up 
a smaller. The disparity could therefore potentially be explained by the minimal impact
of variance in these cities on the overall compactness score, which in turn affects 
optimization and thresholds in recombination. However, further study in this area 
is once again required.

\section{Future Analysis}

The calculation of this measure may be an important indicator of how state geometry and population
distributions affect redistricting. On its own, the average of similarity scores approximately shows what portion of a
reasonable redistricting is already predetermined geometrically.
On the level of individual pairs of plans, the measure provides a mechanism for 
cleverly assigning consistent district identifiers, which can then allow further
comparison in other areas, such as shifts in electoral outcomes across districts 
modified by the redistricting process.

In future analysis, we plan to explore how similarity score correlates with gaps in representation between the popular vote and the elected delegation, and to further study what variables determine the score in the first place (for example, compactness and the number of districts). 
Additionally, the sample size of plans ($N=50$) should be increased significantly to the 
order of thousands of plans, and should be sampled from a first-cut recombination distribution
more in line with previous literature.
Finally, the computational efficiency of this new similarity measure demonstrates its potential as a term in 
an objective function for optimal redistricting algorithms, and we will explore this use case further in future work

\appendix
\printbibliography

\end{document}